\documentclass[letter]{aa}
\usepackage{graphicx}
\usepackage{natbib}
\usepackage{txfonts}
\begin{document}
   \title{The properties of penumbral microjets inclination}

   \author{J. Jur\v{c}\'{a}k
          \inst{1,2}
          \and
          Y. Katsukawa\inst{1}}

   \institute{National Astronomical Observatory of Japan, 2-21-1 Osawa,
Mitaka, Tokyo 181-8588, Japan
        \and
        Astronomical Institute of the Academy of Sciences, Fricova
298, 25165 Ond\v{r}ejov, Czech Republic}

   \date{Received September 15, 1996; accepted March 16, 1997}


  \abstract
   {}
   {We investigate the dependence of penumbral microjets inclination on the
   position within penumbra.}
   {The high cadence observations taken on 10 November 2006 with the Hinode satellite
   through the \ion{Ca}{ii}~H and G--band filters were analysed to determine the inclination of
   penumbral microjets. The results were then compared with the inclination of the magnetic field determined
   through the inversion of the spectropolarimetric observations of the same
   region.}
   {The penumbral microjet inclination is increasing towards the outer edge of the penumbra.
   The results suggest that the penumbral microjet follows
   the opening magnetic field lines of a vertical flux tube that creates the sunspot.}
   {}


   \maketitle
%

\section{Introduction}

\citet[][hereafter Paper~I]{Katsukawa:2007} reported on the existence of small
jet-like features that are observed at penumbral chromospheric layers. The
authors used the term penumbral microjets (PJ) and this nomenclature is also
used in this Letter. These new penumbral phenomena were found using the
observations taken with the Hinode satellite through the broadband
\ion{Ca}{ii}~H filter.

As summarised in Paper~I, the PJs are highly transient events with lifetimes up
to two minutes and lengths of a few thousand kilometres. The width of PJs is
around 400~km for the largest ones where the smallest events are at the
resolution limit of the observations. It can be expected that there are even
smaller undetected penumbral microjets.

As explained in Paper~I, the three-dimensional configuration of the PJs can be
estimated from the different visibilities of these events depending on the
position on the solar disc. The intensity of microjets is comparable to the
intensity of underlaying penumbral filaments, and the PJs can hardly be
identified on observations taken close to the disc centre (if the
running-difference or the high-pass filter are not applied). It implies that
the azimuthal orientation of PJs is the same as for penumbral filaments, i.e.,
radial. In the case of observations taken farther from the disc centre, the
microjets become visible due to the difference in their inclination compared to
the inclination of penumbral filaments that are nearly horizontal. In Paper~I
the authors estimated the inclinations (angle between the local normal line and
the microjet) to be mostly between 40$^\circ$ and 60$^\circ$.

Although the observations with a cadence of four seconds can be achieved (with
a limited field of view and only for a short time), the observations of
penumbra with a 20~sec cadence are currently the best available measurements
that can be used to study the properties of PJs. Taking the lifetimes of PJs
into account, the available high-cadence observations are not fast enough to
study the photospheric counterparts of the PJs onsets because the original
formation area can become dark in 20~seconds. As already pointed out in
Paper~I, there are some indications that the PJs are related to the penumbral
bright grains observed in the photosphere. However, this topic is not discussed
in this Letter and we concentrate on a detailed analysis of the penumbral
microjets inclination and its dependence on the position within the penumbra.
The results are then compared with the inclination of the photospheric magnetic
field.

\section{Observations and data reduction}

We analysed the same data set as in Paper~I. These measurements were taken on
10 November 2006, and they cover a part of the sunspot in AR10923, which was
located at that time at heliocentric coordinates of 6$^\circ$~S and
50$^\circ$~E. Although there are some other observations of sunspots taken with
a 30~sec cadence far from the disc centre, they are either short in time or the
penumbra is of small size so the statistical sample of detected PJs is too
small for the purpose of our study.

\begin{figure}[!b]
  \centering
  \includegraphics[width=\linewidth]{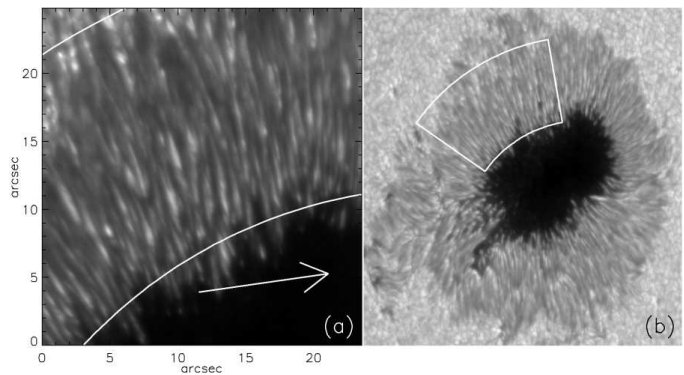}
  \caption{The G--band image of the analysed area (a), where the lines mark
  the position of umbral/penumbral and penumbral/quiet sun boundary and the arrow
  points to the disc centre. The map of continuum intensity of the whole sunspot
  in AR10923 was reconstructed from the SP raster scan (b); see the text for details.}
  \label{overview}
\end{figure}

\begin{figure*}[!t]
  \centering
  \includegraphics[width=0.84\linewidth]{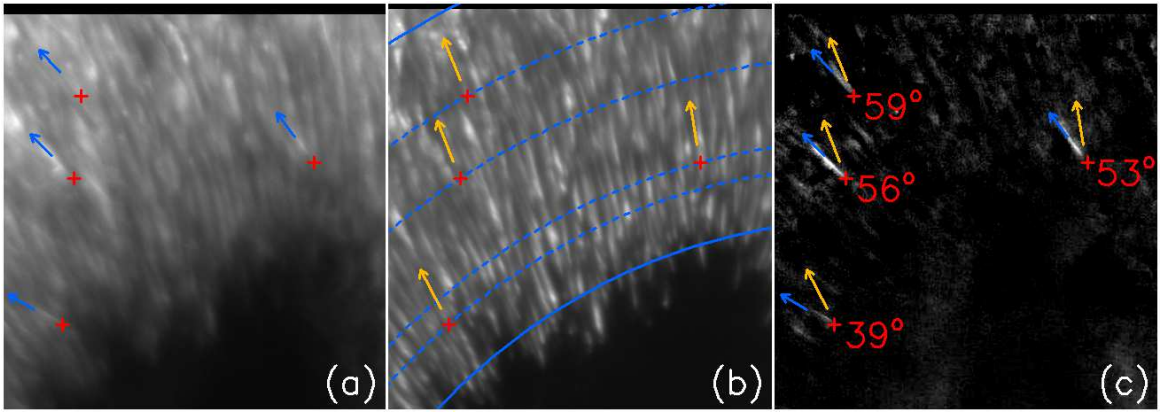}
  \caption{The \ion{Ca}{ii}~H (a), G--band (b), and \ion{Ca}{ii}~H running-difference (c) images
  of the analysed area. The red crosses mark the identified onsets of PJs, the
  blue arrows indicate their orientation, and the orange arrows point in the
  direction of penumbral filaments. The blue lines show the inner and outer
  edges of the penumbra (solid) and the position of PJs (dashed). The resulting values of LRF
  inclinations are shown in the (c) image.}
  \label{jets}
\end{figure*}

The data were taken using the Solar Optical Telescope
\citep[SOT,][]{Tsuneta:2008} onboard Hinode satellite \citep{Kosugi:2007}. The
measurements are unaffected by atmospheric seeing, and the spatial resolution
reaches the diffraction limit of the 50~cm telescope, i.e., 0.2$``$ (150~km)
for the filtergram (FG) data and 0.32$``$ for spectropolarimetric (SP) data.

The analysed FG data were obtained through two broadband filters, the
\ion{Ca}{ii}~H (centred at 396.9~nm with bandwidth of 0.3~nm) and G--band
filter (centred at 430.5~nm with bandwidth of 0.8~nm). The observations took
place between 12:15 and 13:59 UT and are composed of 209 consecutive images.
The data were calibrated with standard routines available under Hinode
Solarsoft. The images in the sequence were aligned to compensate for the drift
of the correlation tracker. The G--band and \ion{Ca}{ii}~H images were
carefully aligned. Figure~\ref{overview}a shows the first G--band image of the
sequence with the estimated position of the umbral/penumbral and
penumbral/quiet sun boundary.

To determine the orientation of the magnetic field, we used the data observed
by the Hinode SP. The measurements were taken in so-called normal mode; the
width of the slit is equivalent to 0.16$``$ and comparable to the scanning
step; the exposure time for one slit position is 4.8~s and results in a noise
level of 10$^{-3} I_c$. More detailed information about this SP observation
mode and the reduction of SP data can be found in \citet{Jurcak:2007}. The
necessary calibration of wavelengths and the normalisation of the observed
Stokes profiles to the continuum intensity of the Harvard Smithsonian reference
atmosphere is also described there. Figure~\ref{overview}b shows part of the
continuum intensity map reconstructed from the SP raster scan of AR10923. The
framed part of the penumbra roughly corresponds to the area observed with FG.
The raster scan of the marked region was taken between 16:25 and 16:40~UT,
approximately 3~hours later than FG measurements.

\section{Estimation of position and inclination}

The alignment between FG and SP data cannot be done precisely since the
penumbra was slowly growing in the meantime between the observations. The
alignment of the inner and outer penumbral boundaries in FG and SP data is
sufficient for the purpose of our study since we do not take the azimuthal
positions into account and the possible inaccuracies in the estimated radial
positions would not affect the results.

The penumbral boundaries are approximately fitted by concentric arcs as shown
in Fig.~\ref{overview}a for FG data and in Fig.~\ref{overview}b for SP data.
The position of PJ in the penumbra is given by the radius of an arc that
crosses the detected onset of the microjet as shown in Fig.~\ref{jets}b.

\subsection{Penumbral microjet inclination}

To compute the inclination angle of the PJs, we first manually determined the
orientation of the microjet either directly from \ion{Ca}{ii}~H data or using
running-difference images. Figures~\ref{jets}a and~\ref{jets}c show examples of
four PJs. The orientations of penumbral filaments are determined from G--band
images, Fig.~\ref{jets}b. Knowing the heliocentric angle and the orientation of
the symmetry line (that connects the disc centre and the microjet onset
position) for each analysed PJ, we can use the following equation
\citep[][]{Muller:2002}
\begin{equation}
\phi=\arctan\left(  \frac{\sin \gamma' \sin \phi'}{\cos \gamma' \sin \theta +
\sin \gamma' \cos \phi' \cos \theta} \right), \label{eq}
\end{equation}
where $\theta$ is the heliocentric angle, $\phi$ the azimuth angle in
line-of-sight (LOS) frame, and $\phi'$ and $\gamma'$ represent the azimuth and
inclination in the local reference frame (LRF) that is defined by the local
normal line (\textit{z} axis) and the orientation of the symmetry line
(\textit{x} axis).

Our final goal is to determine the LRF inclination of the PJ ($\gamma'_{PJ}$).
From the observations we know the heliocentric angle ($\theta$, around
51$^\circ$) and the LOS azimuths of microjet ($\phi_{PJ}$, the angle between
the symmetry line and the PJ) and filament ($\phi_{F}$, the angle between the
symmetry line and the penumbral filament). As shown in Paper~I, we can suppose
that the azimuthal orientation of the PJs is the same as for the penumbral
filaments in the LRF ($\phi'_{PJ}=\phi'_{F}$).

To determine the $\phi'_{F}$ and thus also $\phi'_{PJ}$, we need to estimate
the elevation angle of the penumbral filaments. They are not exactly horizontal
since the Wilson depression must be compensated for in the penumbra. The exact
value of the filaments elevation angle is unknown. We estimate it to be
5$^\circ$ with respect to the local horizontal line. This value corresponds to
the difference found between the filament orientation and the magnetic field
azimuth \citep{Lites:1990a, cwp:2001} that can be caused by the elevation of
penumbral filaments, although there are other possible scenarios for this
difference.

The assumption of a 5$^\circ$ elevation angle gives us the inclination of
penumbral filament in LRF ($\gamma'_F$) as 85$^\circ$. Thus, we know $\theta$,
$\gamma'_F$, and $\phi_{F}$, and Eq.~\ref{eq} can be used to derive the LRF
azimuth value $\phi'_{F}$ that also represents the LRF azimuth of the PJ
($\phi'_{PJ}$). Knowing $\theta$, $\phi'_{PJ}$, and $\phi_{PJ}$, Eq.~\ref{eq}
can be used again to derive the value of $\gamma'_{PJ}$ that represents the
inclination of the microjet in the LRF. In Fig.~\ref{jets}c the derived values
are shown for each of the four depicted jets.

We estimate the error of individually determined inclination values to be up to
10$^\circ$. This uncertainty comes from the manual estimation of the PJ and
filament orientation and becomes even greater in the outer penumbra where the
filament direction could be difficult to estimate from G--band images, and the
microjets are less apparent in \ion{Ca}{ii}~H data due to the increased
intensity of underlying material. Another source of uncertainty is the
incorrect estimation of the onset of PJs. This might result in selecting the
wrong filament in the G--band image.

The unknown value of the penumbral filament elevation, and its possible
dependence on position within the penumbra causes a further increase in the
error. We estimate this error to be in the order of a few degrees. No strong
dependence of this value on the position within the penumbra is expected, as
the absolute values of the elevation angle are expected to be small. Therefore,
the change in PJs inclinations across the penumbra is not significantly
influenced.

\begin{figure}[!t]
  \centering
  \includegraphics[width=\linewidth]{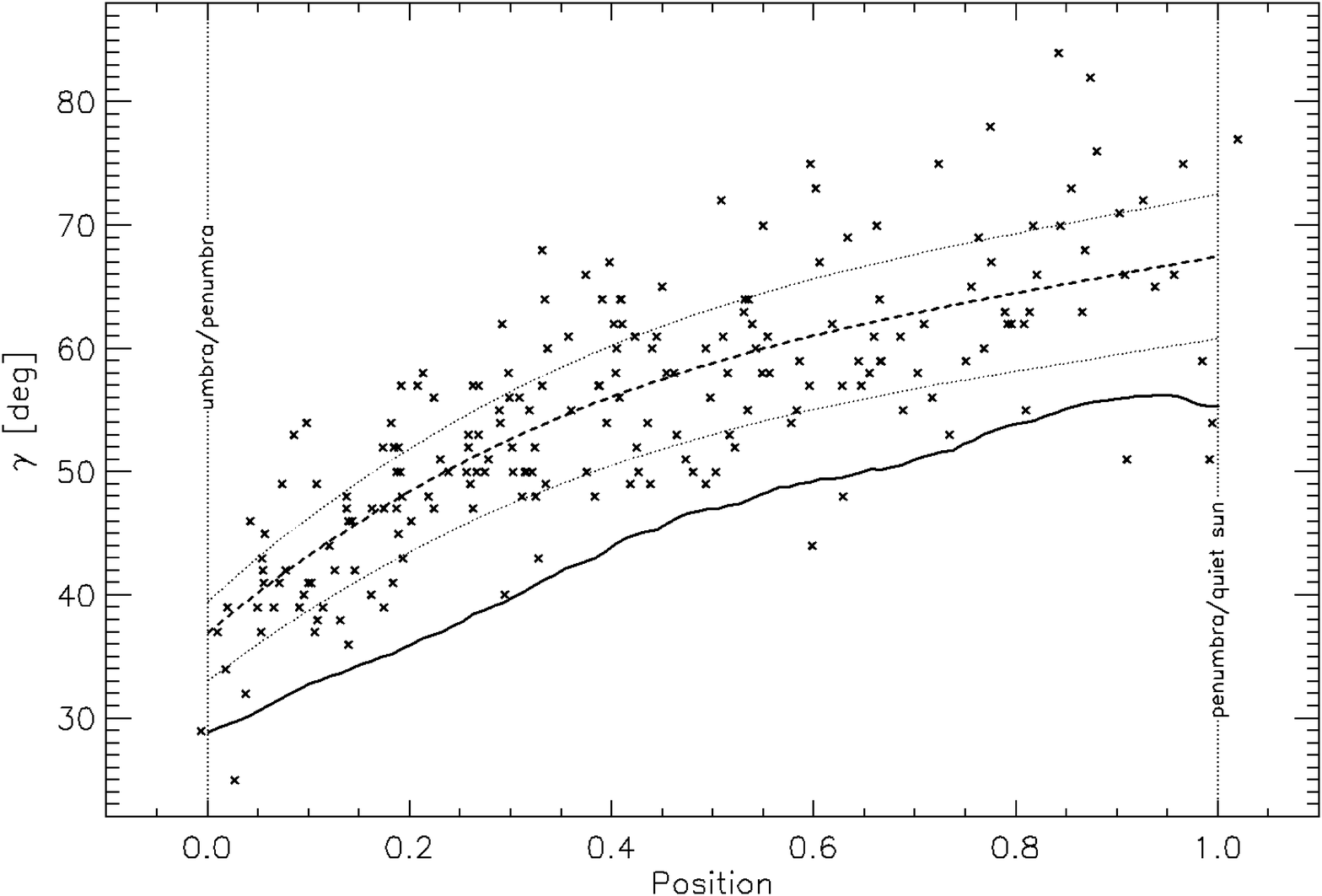}
  \caption{The plot shows the dependence of microjet inclination on the
  position within the penumbra. The 209 detected PJs are represented by $\times$
  symbols. The dashed line is a third-order polynomial fit of the observed
  distribution of microjet inclination through the penumbra, where the dotted
  lines are the alternatives obtained for different values of penumbral
  filament elevation; 0$^\circ$ and 8$^\circ$ for the lower and upper dotted
  line, respectively. The solid line shows the inclination of magnetic field
  at high photospheric layers. The position of umbra/penumbra and penumbra/quiet
  sun boundary are taken as 0 and 1, respectively.}
  \label{plot}
\end{figure}

\subsection{Magnetic field inclination}
\label{mfi}

The Stokes profiles observed at pixels in the marked area in
Fig.~\ref{overview}b are inverted using the SIR code \citep[Stokes Inversion
based on Response functions;][]{Cobo:1992}. Taking the high spatial resolution
of the Hinode SP measurements into account, we use only a one-component
atmospheric model. However, given the observed asymmetries of Stokes profiles,
we allow for the changes in plasma parameters with height in the atmosphere.

The derived values of magnetic field inclination and azimuth are evaluated with
respect to the LOS. After the transformation to the LRF, we determine the
single value of magnetic field inclination at each pixel. As the PJs take place
in the chromosphere, we want to determine the magnetic field inclination at the
highest possible layers of photosphere. According to \citet{cabrera:2005}, the
observed pair of iron lines is the most sensitive to plasma parameters in the
range of optical depths between $\log (\tau)=-1$ and $-2$. As the height in the
atmosphere increases with decreasing $\log (\tau)$, we use the average value of
inclination from the range of $\log (\tau)= \langle-1.5,-2\rangle$. Azimuthal
averages over pixels with the same position in the penumbra are computed to
obtain a curve that shows the magnetic field inclination as a function of the
radial distance in the penumbra.

The obtained curve (solid line in Fig.~\ref{plot}) is comparable in absolute values to the
inclinations of the so-called background component of the penumbral atmosphere
obtained from two-component inversions \citep{Bellot:2004, Borrero:2006}. The
concept of two-components comes from the uncombed model of penumbral atmosphere
\citep{Solanki:1993,vmp:2000}, where the background component has a stronger
and more vertical magnetic field compared to the second component and
represents the properties of a vertical flux tube that creates the sunspot. The
obtained similarity between the inclination values at higher photospheric
layers and the background component inclination confirms that the horizontal
fields presented in the penumbra are restricted to the lower part of the
line-forming region \citep{Bellot:2006, Jurcak:2007}. Our findings can be also
understood in terms of the cusp model introduced by \citet{Scharmer:2006} where
the horizontal fields are restricted to the lowest photosphere, and the
magnetic field at the higher layers is also that of the vertical flux tube.

\section{Comparison of inclination}

In Fig.~\ref{plot}, the inclination of 209 individual PJs identified at various
positions within the penumbra ($\times$ symbols) are shown and the mean
behaviour of PJs inclination (dashed line) is compared with the radial
variation of the magnetic field inclination (solid line). It appears that there
is an almost constant difference between the mean inclination of chromospheric
PJs and the magnetic field in the photosphere, the former one being more
horizontal by about 10$^\circ$, that is, if we take the 5$^\circ$ elevation
angle of penumbral filaments into account. Even if we suppose that the
filaments are exactly horizontal (lower dotted line), the PJs are still more
inclined than the magnetic field in the photosphere.

If we assume that these values of magnetic field inclination describe the
orientation of the field lines of an vertical flux tube that creates the
sunspot, then there is an apparent explanation for the more horizontal
direction of PJs. Such a vertical flux tube has to be opening with height due
to the pressure balance. If we follow one specific magnetic field line then
this becomes more horizontal with height in the atmosphere. This can explain
the relation and also the observed difference between the chromospheric
microjets and the photospheric magnetic field. A microjet following an opening
magnetic field line would appear more horizontal than this same field line at
photospheric layers.

These assumptions also imply that the PJs would become more horizontal with
height; i.e., the observed microjets should not be straight but instead bend
towards the horizontal direction. There are only a few observed cases of PJs in
the analysed data set that appear to be curved. We suppose that the rarity of
curved PJs is mainly caused by the shortness of these events and by the
projection effect to the LOS frame.

\begin{figure}[!t]
  \centering
  \includegraphics[width=0.95\linewidth]{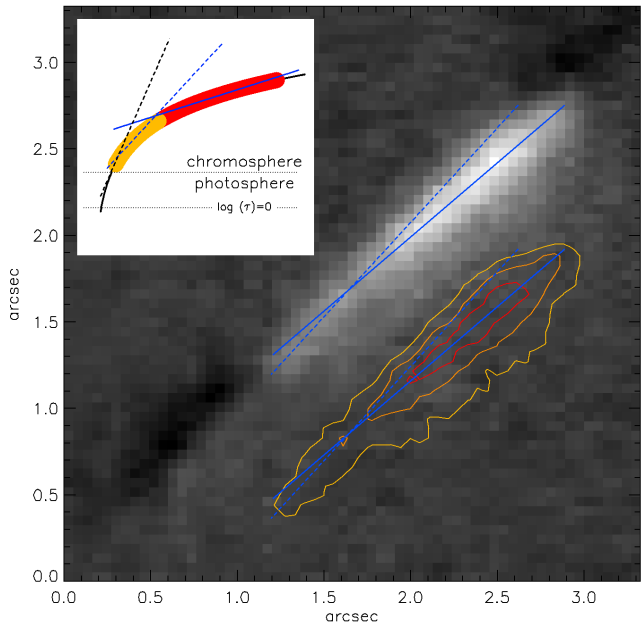}
  \caption{The magnified running-difference image of one of the microjet that
  is slightly curved. The solid blue line is the linear fit of the brightest area of
  the microjet and the dashed blue line represent the initial orientation of the microjet.
  The intensity contours, drawn at an offset of -0.8$``$ along the y-axis are shown
  to emphasise the curvature of the PJ. A sketch of a penumbral microjet is
  shown in the upper left corner. The dashed and solid blue lines represent the orientation
  in the initial and later phase, respectively. The sketched PJ follows the opening field line marked
  by the solid black line and the dashed black line shows its inclination at
  higher photospheric layers.}
  \label{curved}
\end{figure}

Figure~\ref{curved} shows the running-difference image of a microjet (bright
area, detected onset is at lower left part of the image) that appears to be
bending towards the horizontal line with height in the atmosphere. A simplified
sketch of a microjet is plotted in the upper left corner of Fig.~\ref{curved}.
There is a small change of PJ orientation between the faint part (initial
phase, fitted by the dashed blue line) and the bright part (later phase, fitted
by the solid blue line). In the case of the observed PJ, the angle between
these lines is 7$^\circ$ in the LOS frame and corresponds to the difference of
9$^\circ$ in the LRF. This value is comparable to the average difference
between the dashed and solid lines in Fig.~\ref{plot}; i.e., the initial
inclination of this microjet is close to the magnetic field inclination at
higher photospheric layers. From the simplified sketch it is also apparent,
that the sooner the PJ is identified, the closer its inclination to the
magnetic field inclination obtained at higher photospheric layers will be.

The results shown in Fig.~\ref{plot} also imply that the field is becoming more
horizontal with height along the local normal line anywhere in the penumbra.
This does not agree with a simple conception of an opening flux tube that
would result in approximately constant inclination values along the local
normal line. Constant inclination with height was also found by
\citet{Orozco:2005}, who used the inversion of spectral lines to determine the
magnetic field orientation in photosphere and chromosphere. However, the
authors assume that the magnetic field inclination is constant with height in
the photosphere. Thus, the obtained photospheric inclinations might by
influenced by the horizontal component of the magnetic field and are indeed
higher than those obtained in this analysis or from the two-component
inversions. The magnetic field inclinations at the chromospheric layers found
by \citet{Orozco:2005} are comparable to the PJ inclinations reported in this
Letter.

\section{Conclusions}

We identified 209 penumbral microjets in almost two-hour long observations of
the penumbra in AR10923 using the \ion{Ca}{ii}~H images taken with 30~s
cadence. In combination with simultaneous G--band observations, the
inclinations of these microjets (angle between the PJ and local normal line)
are determined along with their approximate position in the penumbra. The
results show a clear increase in the PJ inclination toward the outer penumbral
edge. We find on average inclinations around 35$^\circ$ at the umbra/penumbra
boundary and 70$^\circ$ at the penumbra/quiet sun boundary.

The found radial variation in the PJ inclination resembles the change in
magnetic field inclination at higher photospheric layers across the penumbra.
We find the difference of 10$^\circ$ between the inclinations of magnetic field
lines and penumbral microjets, with the former more vertical. This difference
can be explained easily if we suppose that the PJs follow the magnetic field
lines that are opening with height. There are a few observed PJs that show the
change in inclination with height in the atmosphere and support this
hypothesis.

The observed difference also implies that on average we detect the PJs at
higher atmospheric layers and only rarely at the initial stage when the PJs
inclination angles should be close to (or same as) the magnetic field
inclination at higher photospheric layers. The scatter of individually
determined values of PJs inclination in Fig.~\ref{plot} can be influenced by
the different time spans between the onset and the determination of the PJs
inclination.

Although our results cannot clarify the mechanism responsible for the formation
of the PJs, they imply that the PJs are guided by magnetic field lines that are
fanning out with height. Higher cadence observations of \ion{Ca}{ii}~H and
G--band filtergrams and simultaneous SP measurements are needed to study this
problem in detail.

\begin{acknowledgements} We thank Luis Bellot Rubio and Rolf Schlichenmaier for
helpful suggestions and comments. This work was enabled thanks to the funding
provided by the Japan Society for the Promotion of Science. Financial support
from GA AS CR IAA30030808 is gratefully acknowledged. Hinode is a Japanese
mission developed and launched by ISAS/JAXA, with NAOJ as domestic partner and
NASA and STFC (UK) as international partners. It is operated by these agencies
in cooperation with ESA and NSC (Norway). The computations were partly carried
out at the NAOJ Hinode Science Center, which is supported by the Grant-in-Aid
for Creative Scientic Research The Basic Study of Space Weather Prediction from
MEXT, Japan (Head Investigator: K. Shibata), generous donations from Sun
Microsystems, and NAOJ internal funding.
\end{acknowledgements}

\end{document}